\documentclass[preprint,prl,superscriptaddress,prbib,11pt]{revtex4}
\usepackage{graphicx}
\usepackage{bm}
\usepackage{hyperref}
\usepackage{todonotes}
\usepackage{nicefrac}
\usepackage{bbm}
\usepackage[normalem]{ulem}
\usepackage{booktabs}
\usepackage{subfigure}
\usepackage{rotating}
\usepackage{color}
\usepackage{float}
\usepackage{ragged2e}
\usepackage{amsmath}
\usepackage[switch,columnwise]{lineno}
\begin{document}

\title{Local spectroscopy of loop current order with individual magnetic atoms}
\author{Jiangchang Zheng}
\thanks{These authors contributed equally.}
\affiliation{Department of Physics, The Hong Kong University of Science and Technology, Clear Water Bay, Kowloon, Hong Kong SAR}

\author{Caiyun Chen}
\thanks{These authors contributed equally.}
\affiliation{Department of Physics, The Hong Kong University of Science and Technology, Clear Water Bay, Kowloon, Hong Kong SAR}

\author{Xu Zhang}
\thanks{These authors contributed equally.}
\affiliation{Department of Physics and HK Institute of Quantum Science \& Technology, The University of Hong Kong, Pokfulam Road, Hong Kong SAR}

\author{Daniel J. Schultz}
\thanks{These authors contributed equally.}
\affiliation{Institute for Theoretical Condensed Matter Physics, Karlsruhe Institute of Technology, Wolfgang Gaede Strasse 1, 76131 Karlsruhe, Germany}

\author{Gaopei Pan}
\thanks{These authors contributed equally.}
\affiliation{
Institut f\"ur Theoretische Physik und Astrophysik and W\"urzburg-Dresden Cluster of Excellence ct.qmat,
Universit\"at W\"urzburg, 97074 W\"urzburg, Germany}
\affiliation{Department of Physics and HK Institute of Quantum Science \& Technology, The University of Hong Kong, Pokfulam Road, Hong Kong SAR}

\author{Chen Chen}
\affiliation{Department of Physics, The Hong Kong University of Science and Technology, Clear Water Bay, Kowloon, Hong Kong SAR}

\author{Yuan Da Liao}
\affiliation{Department of Physics and HK Institute of Quantum Science \& Technology, The University of Hong Kong, Pokfulam Road, Hong Kong SAR}

\author{Ganesh Pokharel}
\affiliation{Materials Department, University of California Santa Barbara, California 93106, USA}
\affiliation{Department of Natural Sciences, University of West Georgia, Carrollton, GA 30118, USA}

\author{Andrea Capa Salinas}
\affiliation{Materials Department, University of California Santa Barbara, California 93106, USA}

\author{Qirong Yao}
\affiliation{Department of Physics, The Hong Kong University of Science and Technology, Clear Water Bay, Kowloon, Hong Kong SAR}

\author{Luanjing Li}
\affiliation{Department of Physics, The Hong Kong University of Science and Technology, Clear Water Bay, Kowloon, Hong Kong SAR}

\author{Yizhou Wei}
\affiliation{Department of Physics, The Hong Kong University of Science and Technology, Clear Water Bay, Kowloon, Hong Kong SAR}

\author{Hoi Chun Po}
\affiliation{Department of Physics, The Hong Kong University of Science and Technology, Clear Water Bay, Kowloon, Hong Kong SAR}

\author{Ding Pan}
\affiliation{Department of Physics, The Hong Kong University of Science and Technology, Clear Water Bay, Kowloon, Hong Kong SAR}
\affiliation{Department of Chemistry, The Hong Kong University of Science and Technology, Clear Water Bay, Kowloon, Hong Kong SAR}

\author{Han-Qing Wu}
\affiliation{Guangdong Provincial Key Laboratory of Magnetoelectric Physics and Devices, School of Physics, Sun Yat-sen University, Guangzhou 510275, China}

\author{Stephen D. Wilson}
\affiliation{Materials Department, University of California Santa Barbara, California 93106, USA} 

\author{J\"org Schmalian}
\email[]{joerg.schmalian@kit.edu}
\affiliation{Institute for Theoretical Condensed Matter Physics, Karlsruhe Institute of Technology, Wolfgang Gaede Strasse 1, 76131 Karlsruhe, Germany}

\author{Zi Yang Meng}
\email[]{zymeng@hku.hk}
\affiliation{Department of Physics and HK Institute of Quantum Science \& Technology, The University of Hong Kong, Pokfulam Road, Hong Kong SAR}

\author{Berthold J\"ack}
\email[]{bjaeck@ust.hk}
\affiliation{Department of Physics, The Hong Kong University of Science and Technology, Clear Water Bay, Kowloon, Hong Kong SAR}

\date{\today}

\begin{abstract}

Hidden ordered states—characterized by order parameters that elude conventional probes—pose a fundamental challenge for their identification in quantum materials. Recent experiments report evidence for time-reversal symmetry breaking orbital magnetic order and anomalous transport signatures in the $2a\times2a$ charge density wave state of the kagome metal CsV$_3$Sb$_5$ at a temperature $T<30\,$K. Theoretical analyses propose that a time-reversal symmetry breaking loop-current order could exist as the ground state of this charge density wave. However, this microscopic interpretation remains debated and experimentally unverified. In this work, we employ individual magnetic atoms as local quantum sensors to examine the quasiparticle excitations of the charge density wave in CsV$_3$Sb$_5$ with the scanning tunneling microscope. Our spectroscopic measurements show that the magnetic moment of Co induces a spatially localized $dI/dV$ peak inside the spectral gap of the charge density wave near the Fermi energy. Conducting temperature-dependent spectroscopy, we find that this spectral feature emerges at $T<30\,$K. By comparing our experimental observations with results of quantum many-body simulations and realistic tight-binding model calculations, we show that this spectroscopic signature can be naturally interpreted as a local flux defect in a loop current ordered state, arising from the Kondo coupling of the magnetic moment of Co with the loop current electrons. The excellent agreement between our experimental and theoretical results suggests the presence of loop-current order in the $2a\times2a$ charge density wave of CsV$_3$Sb$_5$ at $T<30\,$K. Our results provide a microscopic picture to the observation of time-reversal symmetry breaking orbital magnetism and anomalous transport signatures detected in measurements of the macroscopic material properties.

\end{abstract}

\maketitle
A microscopic loop current at the atomic scale carries an orbital magnetic moment, which breaks time-reversal symmetry (TRS). The collective orbital magnetism of loop currents arranged on a crystal lattice has long been proposed as a mechanism to engender topologically non-trivial electronic states in materials that lack local magnetic moments~\cite{haldane1988model, hsu1991two, varma1997non, chakravarty2001hidden}. This approach is distinct from established concepts to realize integer and fractional Chern insulators that rely on doping topological insulators with magnetic atoms~\cite{chang2013experimental, chang2023colloquium} or on the magnetic moments of localized charge carriers occupying electronic flat bands~\cite{serlin2020intrinsic, wu2019topological, ledwith2020fractional, abouelkomsan2020particle, li2021spontaneous, devakul2021magic,park2023observation, xu2023observation, lu2024fractional}. Hence, microscopic loop currents could impart novel topological phases of matter with potentially unique properties in a broader range of material classes.

This hitherto elusive concept has drawn renewed attention through the discovery of an anomalous $2a\times2a$ charge density wave (CDW) in the kagome lattice materials of the AV$_3$Sb$_5$ (where A is K, Rb or Cs) family~\cite{jiang2021unconventional, zhao2021cascade, mielke2022time, xu2022three, wu2022simultaneous}. Upon formation of a primary $2a\times2a$ CDW state at higher temperatures ($T\approx90\,$K for CsV$_3$Sb$_5$), a variety of experimental studies have reported the detection of additional symmetry-breaking transitions inside the CDW state at lower temperatures~\cite{guo2022switchable, nie2022charge}. In particular, recent results from electric transport, magnetometry, and muon spin rotation measurements on bulk samples collectively suggest the presence of a time-reversal symmetry breaking phase with out-of-plane orbital magnetism at temperatures below $T\approx30\,$K~\cite{guo2022switchable, khasanov2022time, gui2025probing, wang2025long}, even though the material lacks local spin-magnetic moments. Theoretical analyses of a $3Q$ scattering mechanism between electronic states at the $M-$point of the Brillouin zone ~\cite{feng2021chiral, denner2021analysis} propose that charge correlations~\cite{zhao2021cascade, wang2021electronic, li2022discovery} could indeed favor the formation of the sought-after loop current order (LCO) state that spontaneously breaks TRS. However, this interpretation remains heavily debated because the absence of TRS-breaking signatures in other measurements~\cite{saykin2023high, farhang2023unconventional} agrees with analyses that a time-reversal symmetry preserving charge bond order (CBO) state could be the energetically most favorable ground state of the $2a\times2a$ CDW~\cite{park2021electronic}.

From an experimental viewpoint, this controversy is rooted in the difficulty of isolating unambiguous signatures of a static, antiferromagnetic or perfectly compensated orbital-only order parameter with a vanishingly small net magnetization in the limit of zero magnetic field~\cite{kenney2021absence, liege2024search, gui2025probing} using measurements of the macroscopic electric, magnetic, and optical material properties. Neither can loop currents in these systems be imaged directly. The orbital diameter of $\approx5\,$Å is too small to be detected using magnetic-field sensitive nanoscopic imaging methods, such as SQUID-on-tip and scanning NV microscopy~\cite{thiel2019probing, uri2020nanoscale}, and the orbital magnetism cannot be detected using spin-polarized scanning tunneling microscopy (STM) measurements~\cite{li2022no}. Therefore, further insight into the nature of the well-hidden CDW phase at $T<30$ K requires novel experimental approaches.

In this work, we address this challenge and directly probe the quasiparticle excitations of the $2a\times2a$ CDW order parameter in CsV$_{3}$Sb$_{5}$ with STM using individual magnetic Co atoms deposited on the surface as local sensors (as schematically shown in Fig.~\ref{fig:fig1}a)~\cite{pan2000imaging, balatsky2006impurity, jack2020observation}. Using high-resolution spectroscopy with the STM, we find that the magnetic moment of the Co atoms induces a $dI/dV$ peak at low energies inside the spectral gap of the $2a\times2a$ CDW. The spectroscopic signatures of this $dI/dV$ peak are consistent with theoretical expectations for the quasiparticle excitation of a loop current order state, which arises through the Kondo coupling between the local magnetic moment and the loop current electrons.

\section{Results}
\subsection{Charge Density Wave States in C\lowercase{s}V$_{3}$S\lowercase{b}$_{5}$}

\begin{figure}[H]
    \centering
    \includegraphics[width=1\linewidth]{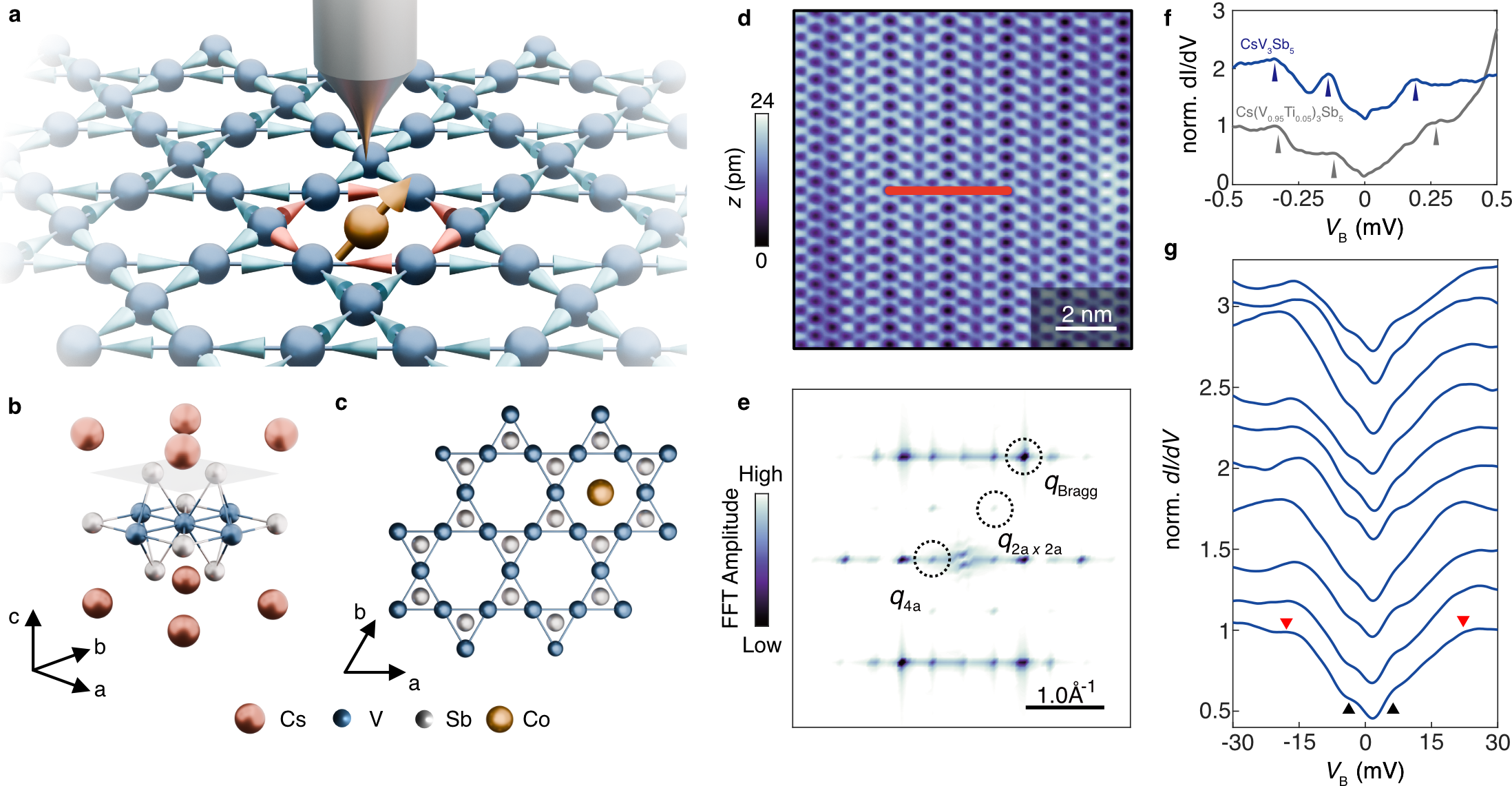}
    \caption{{\bf Quasiparticle spectroscopy of hidden order in CsV$_{3}$Sb$_5$ with individual magnetic atoms} {\bf a,} Schematic presentation of the experimental methodology. Shown is the local response of the local current pattern (red color) in the loop current ordered phase (blue arrows) on the kagome lattice (blue spheres) to the presence of the magnetic moment of a Co atom (orange sphere with arrow). This local perturbation of the loop current order results in a local quasiparticle excitation in the low-energy density of states that can be detected with the tip of a scanning tunneling microscope (STM). {\bf b,} Isometric projection of the unit cell of CsV$_{3}$Sb$_5$. {\bf c,} Lattice structure of CsV$_{3}$Sb$_5$ along the $a-b$ plane, as seen in topographic measurements of the Sb-terminated surface with the STM. The adsorption position of the Co atom in the center of the honeycomb lattice of Sb atoms is shown. {\bf d,} Topography of the Sb-terminated surface of CsV$_{3}$Sb$_5$ recorded with an STM ($V_{\rm B}=-100\,$mV, $I=2\,$nA, $T=4\,$K). {\bf e,} Two-dimensional fast-Fourier transform of the topography shown in panel d. The wave vectors of the Bragg points, as well as the $2a\times2a$ and $4a$ CDWs are indicated. {\bf f,} Normalized differential tunnel conductance ($dI/dV$) spectra recorded on the Sb-terminated surface of CsV$_{3}$Sb$_5$ (solid blue line) and Cs(V$_{0.95}$Ti$_{0.05}$)$_3$Sb$_5$(solid gray line), respectively ($V_{\rm B}=-0.5\,$V, $I=2\,$nA, $V_{\rm m}=1\,$mV, $T=4\,$K). {\bf g,} Shown is a series of $dI/dV$ spectra recorded along the red line in panel d over a smaller bias voltage range ($V_{\rm B}=-30\,$mV, $I=2\,$nA, $V_{\rm m}=1\,$mV, $T=4\,$K). The solid red triangle marks the edges of the V-shaped gap associated with the $2a\times2a$ CDW. and the solid black triangles mark the shoulders of the $4\times1$ CDW gap.}
    \label{fig:fig1}
\end{figure}

The kagome metal CsV$_3$Sb$_5$ crystallizes in a hexagonal structure ($a=b=5.4\,$Å, $c=9\,$Å) and is composed of vanadium (V) antimony (Sb) layers that are stacked with caesium (Cs) along the crystallographic $c$-direction (as seen in Fig.~\ref{fig:fig1}, b and c)~\cite{ortiz2019new}. The V-Sb layer is composed of a kagome lattice of V atoms which coordinates a hexagonal lattice of Sb atoms. Fig.~\ref{fig:fig1}d displays the STM topography recorded on the Sb-terminated surface of a cleaved CsV$_3$Sb$_5$ crystal recorded at a temperature $T=4\,$K. Owing to the comparably weak chemical bond between the Cs and Sb atoms, the cleaved surface is commonly terminated by the Sb layer~\cite{zhao2021cascade}. It has been experimentally shown that CsV$_3$Sb$_5$ hosts a $2a\times2a$ and a unidirectional $4a\times1a$ CDW that arise at temperatures below $\approx90\,$K and $\approx50\,$K, respectively~\cite{zhao2021cascade, jiang2021unconventional, li2022discovery, mielke2022time, xu2022three}. Their presence leads to periodic modulations of the surface topography of CsV$_3$Sb$_5$ that can be detected in STM measurements, as seen in Fig.~\ref{fig:fig1}d~\cite{zhao2021cascade, jiang2021unconventional}. The presence of CDWs can be analyzed by considering the two-dimensional (2D) fast-Fourier transform (FFT) of the surface topography, shown in Fig.~\ref{fig:fig1}e. It features six Bragg peaks ($q_{\rm Bragg}$) that can be associated with the honeycomb lattice of the surface Sb layer. In addition, the 2D-FFT features two more sets of wave vectors, $q_{\rm 2a\times 2a}$ and $q_{\rm 4a}$, which can be associated with the $2a\times2a$ and $4a\times1a$ CDWs. Note that the relative intensity of the $q_{\rm 2a\times 2a}$ peaks in the 2D-FFT is strongly bias voltage-dependent owing to the strong impact of the $q_{\rm 4a}$ CDW on the surface topography~\cite{zhao2021cascade}. Therefore, the relative intensity of the six $q_{\rm 2a\times 2a}$ peaks at $V_{\rm B}=-100\,$mV breaks the six-fold rotation symmetry, as shown in Fig.~\ref{fig:fig1}d.

The presence of the CDWs also strongly affects the electronic states of CsV$_3$Sb$_5$. The electronic band structure around the Fermi energy is dominated by van Hove singularities near the $M-$point in the Brillouin zone. These appear as characteristic peaks in the $dI/dV$ spectrum, as seen in Fig.~\ref{fig:fig1}f~\cite{zhao2021cascade}. Previous angle-resolved photo-electron spectroscopy measurements report that the $2a\times2a$ CDW gaps out parts of the Fermi surface near the $M$-point~\cite{nakayama2021multiple, kang2022twofold}. Consistent with this observation, theoretical models suggest CDW formation via a $3Q$ scattering nesting mechanism~\cite{denner2021analysis, feng2021chiral}. More recent measurements also find that the $4\times1$ CDW arises near the $M'=M/2$-point (equivalently, the new $M$-point of the folded Brillouin zone due to the $2a\times2a$ CDW) at lower energies~\cite{li2023small}. 

Consequently, modifications to the low energy electronic structure near Fermi energy by the CDWs are also reflected in the $dI/dV$ spectra recorded on the surface of CsV$_3$Sb$_5$, as shown in Fig.~\ref{fig:fig1}g. These spectra are dominated by a pronounced V-shaped gap feature with shoulders at $V_{\rm B}\approx\pm20\,$mV (indicated by red solid triangles) that appears on top of a finite $dI/dV$ background. A closer inspection of the $dI/dV$ spectra also reveals the presence of shoulders inside the V-shaped gap at $\approx\pm5\,$mV (indicated by black triangles). The V-shaped gap corresponds to the spectral gap of the $2a\times2a$ CDW~\cite{zhao2021cascade, nakayama2021multiple, wang2021electronic}, and its spectral appearance is consistent with the partial gapping of the Fermi surface via a nesting mechanism. The shoulders detected at $\approx\pm5\,$mV have been previously associated with the $4a\times1a$ CDW~\cite{zhao2021cascade}. Note that these spectral modifications to the density of states near Fermi energy appear on top of a finite $dI/dV$ background, which can be ascribed to electronic bands near the $\Gamma-$ and $K-$point that are unaffected by the presence of the $2a\times2a$ CDW~\cite{kang2022twofold}.

The remainder of this work will focus on the properties of the hidden order appearing below 30\,K. Using high-resolution scanning tunneling spectroscopy (STS) measurements, we will examine the response of the CDW phase to the presence of the atomic-scale magnetic moment of individual Co atoms. We will interpret our experimental observations in terms of excitations of a loop current order parameter, indicating the hidden order as LCO.

\subsection{Detecting the Kondo effect of individual Co atoms at the surface of C\lowercase{s}(V$_{0.95}$T\lowercase{i}$_{0.05}$)$_3$S\lowercase{b}$_5$}

\begin{figure}[H]
    \centering
    \includegraphics[width=1\linewidth]{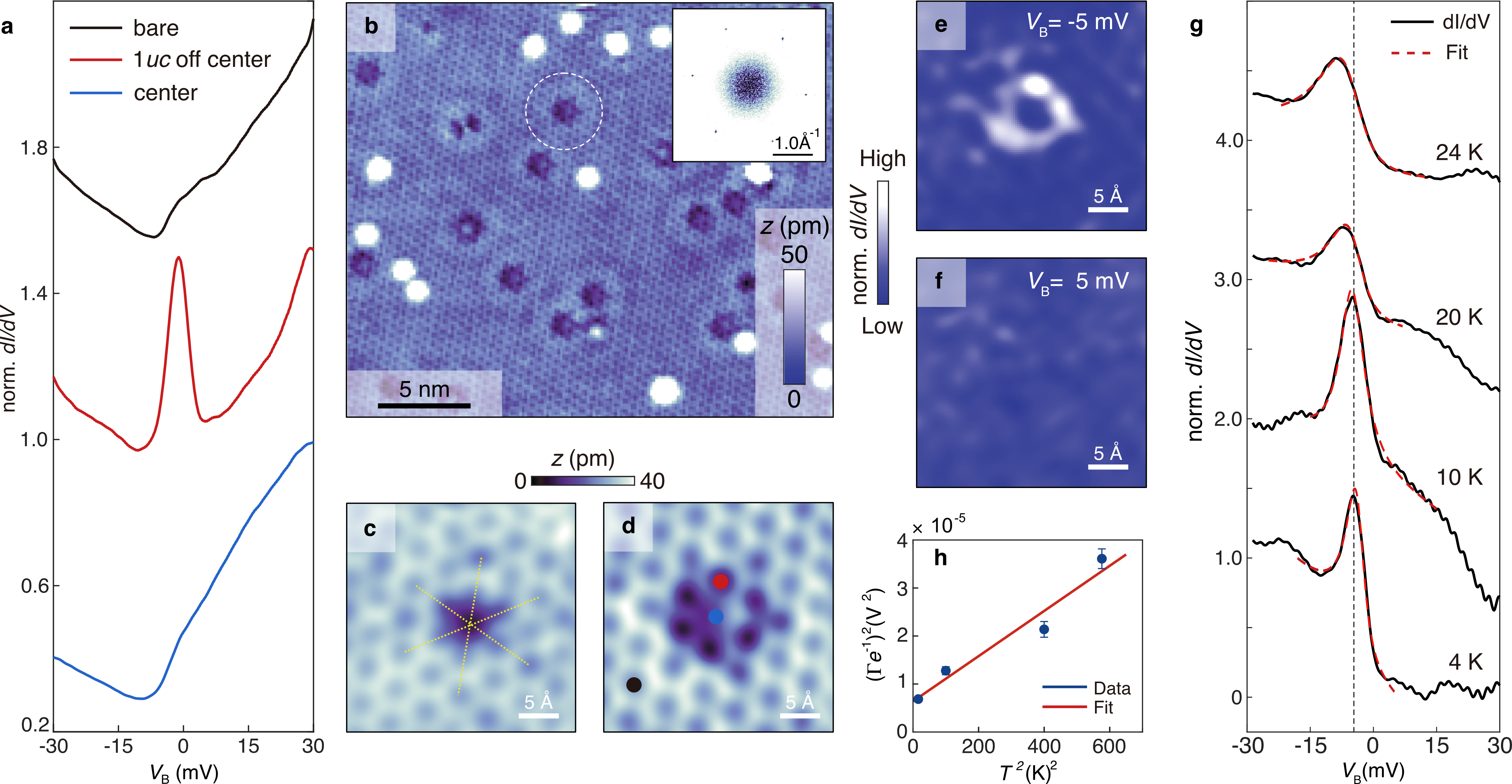}
    \caption{{\bf Kondo effect of individual Co atoms at the surface of Cs(V$_{0.95}$Ti$_{0.05}$)$_3$Sb$_5$.} {\bf a,} Normalized $dI/dV$ spectra recorded on the bare Cs(V$_{0.95}$Ti$_{0.05}$)$_3$Sb$_5$ surface, one unit cell ({\em uc}) away from the Co atom position, and centered at the Co atom position. The STM tip position is indicated by color-coded dots in panel d. The spectra are offset by 0.5 for clarity ($V_{\rm B}=-30\,$mV ,$I=3\,$nA, $V_{\rm m}=1\,$mV, $T=4\,$K). {\bf b,} STM topography of the Sb-terminated surface of Cs(V$_{0.95}$Ti$_{0.05}$)$_3$Sb$_5$ after the Co atom deposition ($V_{\rm B}=-0.5\,$V, $I=2\,$nA, $T=4\,$K). One Co atom is highlighted by the white dashed circle. The inset displays the corresponding 2D-FFT where bright (dark) color correspond to high (low) amplitudes. {\bf c} STM topography of a Co atom position. The yellow dashed lines indicate the position of the Co atom the center of the Sb honeycomb lattice ($V_{\rm B}=0.5\,$V, $I=2\,$nA, $T=4\,$K). {\bf d,} STM topography of the Co atom position in panel c recorded at a different bias voltage ($V_{\rm B}=-0.5\,$V, $I=2\,$nA, $T=4\,$K). {\bf e} and {\bf f,} Two-dimensional maps of the $dI/dV$ amplitude recorded in the field of view of panel d at $V_{\rm B}=-5,\,\text{and}\,5\,$mV, respectively ($I=4\,$nA, $V_{\rm m}=1\,$mV, $T=4\,$K). {\bf g,} Normalized $dI/dV$ spectra (solid lines) recorded one unit cell away from the Co atom position at different indicated temperatures for another Co atom ($V_{\rm B}=-30\,$mV, $I=2\,$nA, $V_{\rm m}=1\,$mV). All spectra were measured on the exact same Co atom and the background $dI/dV$ spectrum recorded on the bare surface was subtracted. The shown spectra are offset by unity and Fano line shape fits are shown as dashed lines. The vertical black line delineates the $dI/dV$ peak position. The comparability of the temperature-dependent $dI/dV$ spectra recorded at different temperatures is maintained by ensuring the absence of tip changes through slow temperature change rates and spectroscopic control measurements (not shown). Details of the background subtraction and fitting procedure are shown in Sec.~C.2 of the suppl.~materials. {\bf h,} Displayed is the dependence of the Fano line width $\Gamma$, as extracted from the least-square fits to the $dI/dV$ spectra shown in panel g, on the temperature $T$, as well as a linear least-square fit to the data points (red line). $e$ denotes the electron charge.}
    \label{fig:fig2}
\end{figure}

To understand the interaction of the $2a\times2a$ CDW with a local magnetic moment, we first establish the response of the itinerant electronic states of CsV$_{3}$Sb$_{5}$ to the presence of a local magnetic moment, when charge correlations are absent. It has been reported that titanium doping with $x>0.05$ suppresses the $2a\times2a$ and $4a\times1a$ CDWs by shifting the van Hove singularities away from the Fermi energy~\cite{liu2023doping}. In this work, we examine Cs(V$_{0.95}$Ti$_{0.05}$)$_3$Sb$_5$ crystals with $x=0.05$ Ti doping. In agreement with previous results~\cite{huang2025revealing}, our analysis of the $dI/dV$ spectrum over a wider bias voltage range (see Fig.~\ref{fig:fig1}f) shows that Ti-doping lowers the Fermi energy by approx. $40\,$meV. A detailed study of the STM topographies recorded on the surface of these samples, presented in Sec.~A of the supp.~materials, confirms the absence of the $2a\times2a$ and $4a\times1a$ CDWs~\cite{yang2022titanium}. Consistent with this observation, $dI/dV$ spectra recorded on this sample surface lack the V-shaped gap and low-energy shoulders at Fermi energy, which were previously detected on undoped CsV$_{3}$Sb$_{5}$, as seen in Fig.~\ref{fig:fig2}a (black line).

Individual cobalt atoms deposited on metallic surfaces are a paradigmatic and well-studied Kondo system, where the direct coupling between the magnetic moment of Co and itinerant states results in pronounced $dI/dV$ peaks near Fermi energy and Kondo temperatures of a few tens of kelvins~\cite{ternes2008spectroscopic}. In our study, we evaporated Co atoms on the cold ($T=4\,$K) surface of Cs(V$_{0.95}$Ti$_{0.05}$)$_3$Sb$_5$ (see Methods section) to prevent cluster formation through thermal diffusion. Figure~\ref{fig:fig2}b shows an STM topography after the deposition of Co atoms. This topography was recorded at a bias voltage $V_{\rm B}=-0.5\,$V in order to suppress topographic features associated with the Ti dopants (see Sec.~A of the supp.~materials for more details). Note the absence of the $q_{\rm 2a\times 2a}$ peaks in the 2D-FFT (inset of Fig.~\ref{fig:fig2}b). Co atoms are distributed across the surface and can be distinguished from other randomly occurring surface defects, such as Cs surface adatoms that appear as white dots~\cite{zhao2021cascade}, by a characteristic star-of-David (SoD) shape. We highlight one Co atom by a white dashed circle in Fig~\ref{fig:fig2}b. We detect small variations in the topographic appearance of the SoD shape that presumably result from variations of the electronic background due to Ti doping. Moreover, we find that the topographic appearance of the Co atoms is strongly bias voltage dependent and evolves from a SoD shape at $V_{\rm B}<0$ to a near point-like appearance at $V_{\rm B}>0$, as seen in Fig.~\ref{fig:fig2}, c and d and Sec.~B of the supp.~materials.

The STM topographies also reveal that the Co atoms do not assume ad-atomic positions on top of the sample surface. Instead, they integrate into the surface layer as indicated by their vanishing apparent topographic height. The analysis of the bias voltage dependent STM topographies indicates that all examined Co atoms reside at the center of the honeycomb lattice of the Sb atoms ({\em c.f.}~Fig.~\ref{fig:fig1}c). This observation is reproduced by {\em ab-initio} simulations of the surface adsorption process of Co atoms. Our simulation results show that Co atoms integrate into the honeycomb center of the topmost Sb layer and form bonds with $d$- and $p$-orbitals of the vanadium and antimony atoms, respectively (details of the model calculation results are discussed in Sec.~B of the supp.~materials).

We have conducted STS measurements to characterize the response of Cs(V$_{0.95}$Ti$_{0.05}$)$_3$Sb$_5$ to the presence of individual Co atoms. $dI/dV$ spectra recorded at, near and away from the Co atom position (as shown in Fig.~\ref{fig:fig2}d) are presented in Fig.~\ref{fig:fig2}b. The $dI/dV$ spectra recorded at and far away from the Co-atom position are comparable and dominated by a reduction of the $dI/dV$ amplitude near the Fermi energy $E_{\rm F}$. Interestingly, the $dI/dV$ spectrum (red solid line) recorded one unit cell away from the Co atom exhibits a pronounced peak at $V_{\rm B}\approx-5\,$mV. The off-center-position of this $dI/dV$ peak can be clearly seen in spectroscopic mapping measurements. $dI/dV$ maps recorded at the bias voltage of this peak exhibit a circular enhancement of the $dI/dV$ amplitude around the Co atom position (as shown in Fig.~\ref{fig:fig2}e). This halo-like enhancement of the $dI/dV$-amplitude is most pronounced at $V_{\rm B}=-5\,$mV and all but absent at $V>0\,$mV (as seen in Fig.~\ref{fig:fig2}f). We show $dI/dV$ maps recorded at other bias voltages, as well as additional data sets that reproduce these results in Sec.~C of the supp.~materials. Note that all Co atoms on the surface of Cs(V$_{0.95}$Ti$_{0.05}$)$_3$Sb$_5$ that we have examined (total number $>20$) exhibit this $dI/dV$ peak at $V_{\rm B}<0$ with halo-like spectral weight distribution despite small variations in the topographic characteristics of the SoD pattern.

These observations suggest the presence of the Kondo effect in Cs(V$_{0.95}$Ti$_{0.05}$)$_3$Sb$_5$, which is induced by the coupling of the magnetic moment of the Co atom to the spins of the itinerant electronic states at Fermi energy. It has been established that a $dI/dV$ peak induced by the Kondo effect can be identified by a unique temperature dependence~\cite{ternes2008spectroscopic}. To verify our hypothesis, we examine the temperature dependence of this $dI/dV$ peak detected near another Co atom, as seen in Fig.~\ref{fig:fig2}g. Increasing the experimental temperature continuously from 4 to 24\,K increases its spectral width $\Gamma$. Our analysis shows that this broadening effect cannot be solely explained by thermal broadening $\Gamma\propto k_{\rm B}T$. Instead, the temperature-dependent $dI/dV$ peak can be better described with a Fano line shape using least-square fits (shown in Fig.~\ref{fig:fig2}g). Details and accuracy of the fitting procedure are described in Sec.~C.2 of the supp.~materials. 

The extracted temperature dependence of $\Gamma$ follows the well-known Kondo expression $\Gamma(T)=\sqrt{\pi(k_{\rm B}T)^2+2(k_{\rm B}T_{\rm K})^2}$ (as shown in Fig.~\ref{fig:fig2}h), which is bounded by the intrinsic spectral width of the Kondo resonance for $T\rightarrow0\,$K ($k_{\rm B}$ and $T_{\rm K}$ denote Boltzman's constant and the Kondo temperature, respectively)~\cite{nagaoka2002temperature}. Moreover, closer inspection of the spectra in Fig.~\ref{fig:fig2}g reveals that the Kondo peak shifts to slightly more negative bias voltages near $T_{\rm K}$, an effect that can possibly be ascribed to the crossover from potential scattering at $T<T_{\rm K}$ to spin-flip scattering $T>T_{\rm K}$, respectively~\cite{nagaoka2002temperature}. We can accurately reproduce this temperature-dependent Fano line shape analysis using data recorded on another Co atom [see Figs.~S7 and S8]. This establishes the observation of the Kondo effect induced by Co atoms deposited on the surface of Cs(V$_{0.95}$Ti$_{0.05}$)$_3$Sb$_5$. Further details of the detected Kondo effect are discussed in Sec.~C.3 of the suppl.~materials. Finally, note that our attempts to manipulate the lateral and vertical position of the Co atoms with the STM tip failed, probably owing to its sub-surface position.

\subsection{Visualizing a localized quasiparticle excitations in the CDW gap of C\lowercase{s}V$_{3}$S\lowercase{b}$_{5}$ induced by a Co atom}

\begin{figure}[H]
    \centering
    \includegraphics[width=1\linewidth]{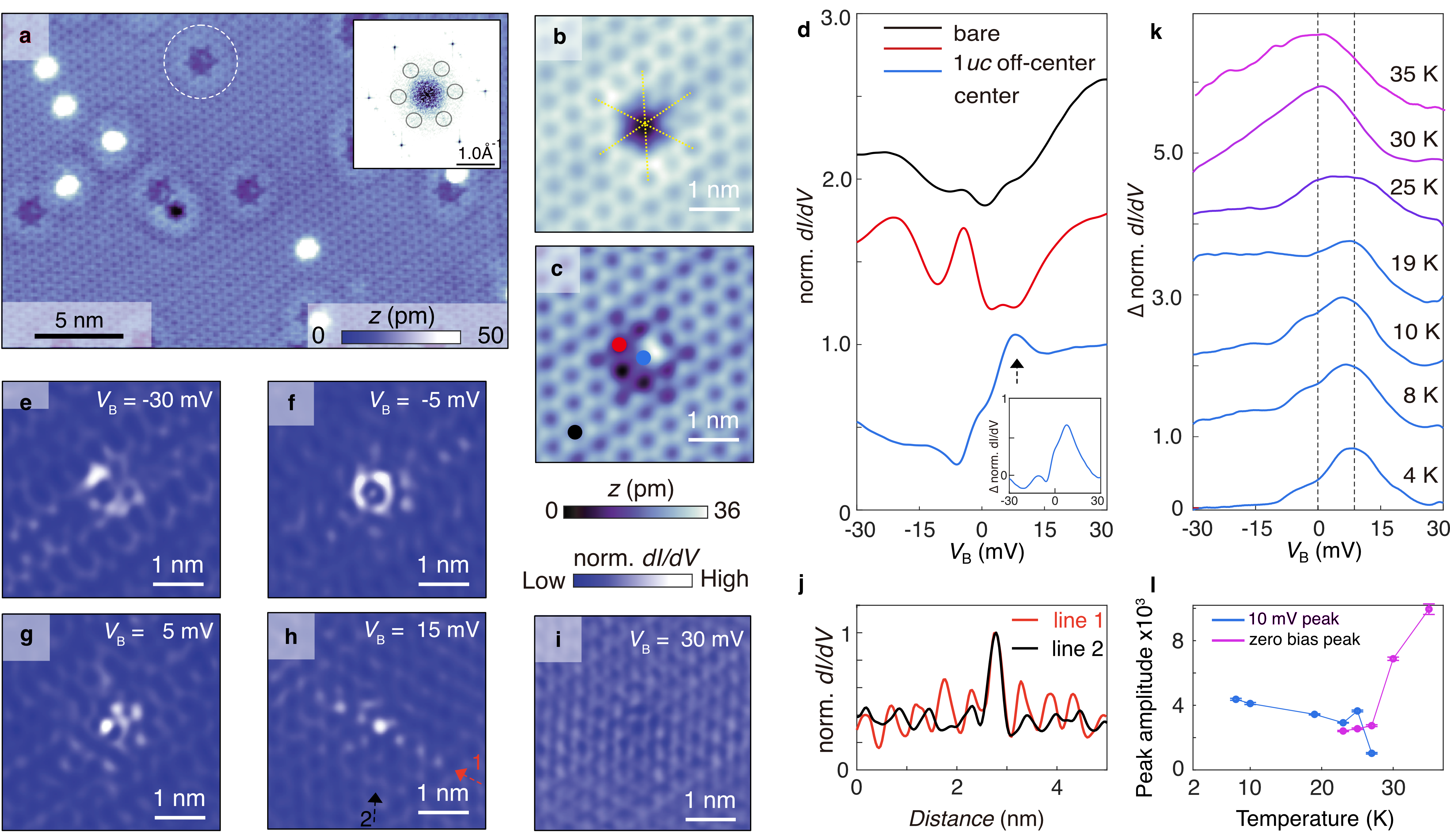}
    \caption{{\bf Visualizing the low-energy quasiparticle excitations of CsV$_{3}$Sb$_5$ at the atomic scale} {\bf a,}  STM topography of the Sb-terminated surface of CsV$_{3}$Sb$_5$ after the Co atom deposition ($V_{\rm B}=-0.5\,$V, $I=2\,$nA, $T=4\,$K). The inset displays the corresponding 2D-FFT where bright (dark) color correspond to high (low) amplitudes. The positions of the $2a\times2a$ quasiparticle interference (QPI) peaks are indicated by black circles. Owing to the bias voltage dependence of the topographic contrast, the QPI peaks have a small amplitude at $V_{\rm B}=-0.5\,$V~\cite{zhao2021cascade}. {\bf b,} STM topography of a Co atom position. The yellow dashed lines indicate the position of the Co atom the center of the Sb honeycomb lattice ($V_{\rm B}=-0.5\,$V, $I=2\,$nA, $T=4\,$K). {\bf c,} STM topography of the Co-atom position in panel b recorded at a different bias voltage ($V_{\rm B}=0.5\,$V, $I=2\,$nA, $T=4\,$K). {\bf d,} Normalized $dI/dV$ spectra recorded on the bare CsV$_3$Sb$_5$ surface, one unit cell ({\em uc}) away from the Co atom position, and centered at the Co atom position. The STM tip position is indicated by color-coded dots in panel c. The spectra are offset by unity ($V_{\rm B}=-30\,$mV, $I=3\,$nA, $V_{\rm m}=1\,$mV, $T=4\,$K). The inset shows the difference spectrum $\Delta\,\text{norm}.\, dI/dV$ as the difference between the normalized $dI/dV$ spectra recorded at the Co atom position (blue dot in panel c) and on the bare surface (black dot in panel c). {\bf e}-{\bf i,} Shown are two-dimensional maps of the $dI/dV$ amplitude recorded in the same field of view as panel c at $V_{\rm B}=-30,\,-5,\,5,\,15,\,\text{and}\,30\,$mV ($I=4\,$nA, $V_{\rm m}=1\,$mV, $T=4\,$K). {\bf j,} Shown is the spatial dependence of the $dI/dV$ amplitude across the Co atom position along two spatial directions. The two directions are indicated by color-coded arrows in panel h. {\bf k,} Shown is the temperature-evolution of the normalized $dI/dV$ spectra recorded at the position of Co atom 3 after subtracting the $dI/dV$ spectrum recorded on the bare substrate ($V_{\rm B}=-30\,$mV, $I=3\,$nA, $V_{\rm m}=1\,$mV). The full data set and background spectra are shown in Ext.~Data.~Fig.~2, a-c. The vertical dashed lines delineate the spectral position of the $dI/dV$ peak at $\approx10\,$mV and zero bias, respectively. {\bf l,} Shown are the fitted amplitudes of the $dI/dV$ peak at $\approx10\,$mV and zero bias voltage as extracted from the curve fitting analysis presented in Sec.~D.3 of the supp.~materials.}
    \label{fig:fig3}
\end{figure}

In the next step, we examine the response of the hidden order in pristine CsV$_{3}$Sb$_{5}$ to the presence of the local magnetic moment of a Co atom. We evaporated Co atoms on the cold surface of undoped CsV$_{3}$Sb$_{5}$ (as shown in Fig.~\ref{fig:fig3}a), which exhibits the characteristic $2a\times2a$ and $4a\times1a$ CDWs, respectively. The Co atoms are randomly distributed across the sample surface. Consistent with our observations on Ti-doped CsV$_{3}$Sb$_{5}$, they exhibit bias-voltage dependent topographic characteristics and the analysis of their spatial position (as shown in Fig.~\ref{fig:fig3}b) demonstrates their integration into the top-most Sb-layer at the honeycomb center position. These observations indicate that Co atoms deposited on the surface of doped and undoped CsV$_{3}$Sb$_{5}$ exhibit the same orbital bond structure.

We have conducted STS measurements to characterize the low-energy electronic density of states of CsV$_3$Sb$_5$ in the presence of individual Co atoms. $dI/dV$ spectra recorded on the sample surface away from the Co atom position (black dot in Fig.~\ref{fig:fig3}c) feature the characteristic V-shaped gap of the $2a\times2a$ CDW, as shown in Fig.~\ref{fig:fig3}d (black line). Spectra recorded one unit cell away from the Co atom position (red dot in Fig.~\ref{fig:fig3}c) additionally exhibit a pronounced $dI/dV$ peak at $V_{\rm B}\approx-5\,$mV (red line in Fig.~\ref{fig:fig3}d), and the spectral weight of this peak is distributed in a halo-shaped structure in real space, as seen in Fig.~\ref{fig:fig3}f). These observations reproduce the results from our measurements on Cs(V$_{0.95}$Ti$_{0.05}$)$_3$Sb$_5$ and indicate the presence of the Kondo effect induced by the magnetic moment of the Co atom in CsV$_{3}$Sb$_{5}$.

Interestingly, when the STM tip is located right at the Co atom position (blue dot in Fig.~\ref{fig:fig3}c), we detect one additional $dI/dV$ peak that appears inside the pseudogap at positive bias voltage $V_{\rm B}~\approx10\,$mV (blue line in Fig.~\ref{fig:fig3}d; the peak is highlighted by a black arrow). This spectral characteristics of this $dI/dV$ peak are particularly apparent in the difference spectrum $\Delta_{\rm norm}\,dI/dV=[dI/dV(\text{Co atom position})-dI/dV(\text{bare surface)}]$ between the spectrum recorded on the Co atom position and on the bare surface (inset of Fig.~\ref{fig:fig3}d). We also examined the real space distribution of this spectral feature. $dI/dV$ maps (shown in Fig.~\ref{fig:fig3}, e-i) recorded in the same field of view as Fig.~\ref{fig:fig3}c establish that the spectral weight of this $dI/dV$ peak is predominantly localized at the honeycomb center of the Sb lattice. This contrasts with the halo-like appearance of the Kondo effect around the Co atom position. Crucially, the spectral weight of this $dI/dV$ peak is strongest at $V_{\rm B}>0\,$mV, and it can only be detected inside the CDW gap at $V<20\,$mV; however, no spectral weight can be detected at energies outside the CDW gap (as shown in Fig.~\ref{fig:fig3}i at $V_{\rm B}=30\,$mV).

We detect a spatially localized $dI/dV$ peak inside the V-shaped spectral gap at $V_{\rm B}>0$ for all Co atoms ($N>50$) examined that were deposited on the surface of three different cleaved CsV$_{3}$Sb$_{5}$ samples). This illustrates the robustness of our experimental observation. Additional data sets are shown in Sec.~D.1 of the supp.~materials. We have also analyzed the profile of the spectral weight distribution of this $dI/dV$ peak in real space. As shown in Fig.~\ref{fig:fig3} j, we find that it extends for more than one nanometer along one direction of an in-plane lattice vector but remains localized to the Co atom position along any other direction. This observation indicates that the underlying electronic state breaks the $C_{6z}$ rotation symmetry of the kagome lattice. We can reproduce the observation of $C_{6z}$-breaking $dI/dV$ peaks in experiments on other Co atoms (see Sect.~D.2 of supplement.~materials), supporting the robustness of this characteristic feature. We also find a subset of Co atoms for which the spatial distribution of the $dI/dV$ peak does not break $C_{6z}$. A detailed analysis of STM topographies (see Sec.~D.2 of the suppl.~materials) reveals that the $C_{6z/2z}$-character of this $dI/dV$ peak is consistent with the local symmetry of the underlying adsorption site of the Co atom within the $2a\times2a$ CDW unit cell, which has either $C_{6z}$ or $C_{2z}$ symmetry~\cite{christensen_loop_2022}.

Finally, we have also characterized the temperature evolution of the low energy density of states to shed light on the origin of this $dI/dV$ peak at $V_{\rm B}\approx10\,$meV. In Fig.~\ref{fig:fig3}k, we plot the difference spectrum $\Delta_{\rm norm}\,dI/dV$ recorded at the center of another Co atom (labeled `Co atom 3') at different temperatures. The corresponding raw data and background spectra recorded on the bare substrate, as well as measurements at additional temperatures, are shown in Ext.Data.Fig.~1,~a-c. As the temperature increases from 4\,K to 23\,K, this $dI/dV$ peak continuously broadens. Above 23\,K, this peak changes its spatial appearance and a distinct $dI/dV$ peak at zero bias voltage arises at $T\geq30\,$K. We have conducted a careful curve fitting analysis to track this temperature evolution, as presented in Sec.~D.3 of the supp.~materials. We find that the $dI/dV$ peak at $V_{\rm B}\approx10\,$mV becomes gradually suppressed with increasing temperature and vanishes at 30\,K, and a second $dI/dV$ peak at zero bias voltage arises at $T\geq23\,$K. This is illustrated by the fitted peak amplitudes shown in Fig.~\ref{fig:fig3}l. Note that the disappearance of the $dI/dV$ peak at 30\,K cannot be caused by spectral broadening $\Delta E$ of the Fermi-Dirac distribution at finite temperature. Our analysis (see Sec.~D.4 of suppl.~materials) shows that the impact of thermal broadening $\Delta E\approx9\,$meV at 30\,K on the peak characteristics is minor owing to its large intrinsic spectral width (approx.~15\,mV at 
4\,K).

On the other hand, $dI/dV$ spectra recorded one unit cell away from the Co atom position, presented in Ext.Data.Fig.~2, show that the Kondo peak at $V_{\rm B}\approx5\,$mV broadens continuously when the temperature increases from 4\,K to 35\,K. Note that small spectral shifts of the Kondo peak above 10\,K can be attributed to temperature-dependent changes of the background density of states ({\em c.f.~}Ext.Data.Fix~1a). Our analyses of the temperature-dependent Kondo linewidth find $T_{\rm K}=29.1\pm0.7\,$K, a 50$\%$ increase in $T_{\rm K}$ compared to the Kondo temperature found for Ti-doped CsV$_3$Sb$_5$. Because $T_{\rm K}$ is exponentially sensitive to the density of states at Fermi energy~\cite{ternes2008spectroscopic}, the $T_{\rm K}$ enhancement could reflect changes to the Fermi surface induced by the $2a\times2a$ CDW through band folding~\cite{li2023small}. This temperature evolution, which we reproduce on another Co atom (labeled `Co atom 4') in Ext.Data.Figs.~1 and 2, establishes a qualitative difference between the $dI/dV$ peak of the conventional Kondo effect detected at $V_{\rm B}\approx-5\,$mV and the additional $dI/dV$ peak detected at $V_{\rm B}\approx10\,$mV. This finding suggests that this spectral feature induced by the Co atom inside the V-shaped gap of the $2a\times2a$ CDW must have a different origin.

\subsection{Response of the CDW to non-magnetic surface adsorbates}

Our experimental observations show that the presence of an individual Co atom induces a spatially localized $dI/dV$ peak at $V_{\rm B}\approx10\,$mV within the V-shaped gap of the $2a\times2a$ CDW. Next, we strengthen the experimental link between the detection of this spectral feature and the presence of a magnetic moment by conducting a set of control experiments using nonmagnetic surface adsorbates. Such control experiments are critical, because spectral in-gap excitations, as detected in spectroscopic measurements with the STM, can generally occur in charge density wave materials~\cite{hildebrand2014doping, lutsyk2023influence} as well as in semiconductors~\cite{richardella2010visualizing, edelberg2019approaching} and topologically trivial and non-trivial insulators~\cite{roman2021band, venkatachalam2011local, lu2011non, luican2014screening, slager2015impurity, feldman2016observation, diop2020impurity} through the Coulomb potential of non-magnetic atomic-scale impurities~\cite{lu2011non, slager2015impurity, diop2020impurity}. The presence of such a potential can localize a simple impurity state that can be detected as an in-gap $dI/dV$ peak that is localized in real space.

To rule out these scenarios, we conducted a set of control experiments that are described in detail in Sec.~E of the supp.~materials. We examined the response of the electronic states of CsV$_{3}$Sb$_{5}$ at low energies to a range of non-magnetic surface impurities. We recorded $dI/dV$ spectra on naturally abundant Cs adatoms and carbon monoxide molecules adsorbed to the sample surface. Moreover, we also examined the effect of individual vanadium atoms that were intentionally deposited on the sample surface (see Methods section). All these impurities are considered to be non-magnetic; the non-magnetic character of the adsorbed vanadium atoms is confirmed through the absence of the Kondo effect. We find that none of these adsorbate species induces a spatially localized $dI/dV$ peak inside the V-shaped gap. Hence, these results indicate that the low-energy $dI/dV$ peak at $V_{\rm B}\approx10\,$mV inside the V-shaped CDW gap is not caused by a local Coulomb potential but rather arises through the coupling between the magnetic moment of the Co atom to the hidden order in the CDW phase.

\section{Discussion}

In summary, our STM experiments show that the magnetic moment of an individual Co atom adsorbed to the CsV$_{3}$Sb$_{5}$ surface induces a $dI/dV$ peak at $V_{\rm B}\approx10\,$mV within the V-shaped gap of the $2a\times2a$ CDW, as seen in Fig.~\ref{fig:fig3}. The unique temperature dependence of this spectral feature, marked by its temperature-induced suppression at $T\approx30\,$K, indicates that it arises within the TRS-breaking phase of the $2a\times2a$ CDW previously detected below this temperature~\cite{khasanov2022time, gui2025probing, wang2025long}. The only known ground state of the $2a\times2a$ CDW that breaks TRS is the LCO state~\cite{denner2021analysis, feng2021chiral, park2021electronic}. Therefore, we will compare our experimental observations with results from quantum many-body simulations and realistic tight-binding model calculations, arguing that our experimental observations below 30\,K are fully consistent with the presence of an LCO state.

\subsection{Kondo coupling of a local magnetic moment to the loop current order}

Our experimental observations indicate that the spatially localized $dI/dV$ peak arises through the local interaction between the atomic-scale magnetic moment of the Co atom and the loop current state. To account for this effect at a theoretical level, we implement a tight-binding model Hamiltonian for the loop current order phase~\cite{feng2021chiral}. A cartoon of the loop current pattern is shown in Fig.~\ref{fig:fig4}(a). Our model captures the interaction between a local spin $S=1/2$ magnetic moment and the spins of the loop current electrons via an antiferromagnetic Kondo coupling of strength $J_{\rm K}$. Model and calculation details are described in Sec.~F of the supp.~materials.

For our calculations, we consider a $S=1/2$ moment at the center of hexagon 4 and solve this model using exact diagonalization (ED) computation on the $2a\times 2a$ unit cell of the CDW. To understand the impact of this Kondo like coupling on the loop current order, we first consider the effective magnetic moment $M_{\text{eff}}=\left\langle\left(\frac{1}{6}\sum_{\langle i,j\rangle,s,s^\prime} \tfrac{1}{2}c_{i, s}^{\dagger} \vec{\sigma}_{s, s^{\prime}} c_{j, s^{\prime}}+\text{H.c.}\right)^2 \right\rangle$ at the hexagon site of the impurity as a function of the Kondo coupling strength $J_{K}$ normalized by the hopping paramter $t$ (Fig.~\ref{fig:fig4}b). The quantity $M_{\text{eff}}$ can be understood as the effective magnetic moment of a spin singlet bond creation at the hexagon of the impurity site which forms owing to the antiferromagnetic Kondo coupling term between the magnetic impurity and conduction electrons.

Our analyses show that at sufficiently strong coupling strength, $J_{\rm K}>2t$, this screening process suppresses spin-conserving hopping along the hexagon sites and reduces the local loop current $j=\sum_{\langle i,j\rangle\in h,s}(-1)^{\eta_{ij}}\,i\,\langle c_{i, s}^{\dagger}c_{j, s}-c_{j, s}^{\dagger}c_{i, s}\rangle$ around the hexagon (the notation $\langle i,j\rangle \in h$ refers to bonds around a hexagon), as shown in Fig.~\ref{fig:fig4}c, an effect that increases with increasing coupling strength $J_{\rm K}>4t$. A perturbation theory analysis of this Kondo interaction, as described in Sec.~G of the supp.~materials, qualitatively reproduces this effect, validating this key result of our numerical many-body simulation. Hence, the local interruption of the LCO through this Kondo spin singlet generation--the local Kondo coupling can be regarded as a competing mechanism--can be interpreted as to induce a flux defect in the periodic flux pattern generated by the LCO. The flux pattern of the unperturbed LCO state is shown in Fig.\ref{fig:fig4}a (see Sec.~I of the supp.~materials for details of the flux pattern construction).

\subsection{Impact of a flux defect on the local electronic density of states}

We now examine the impact of such local suppression of the loop current order parameter on the local electronic density of states by considering the realistic electronic band structure of CsV$_3$Sb$_5$ in a tight-binding model. Previous theoretical studies of LCO focus on a subset of atomic orbitals which are odd under the mirror reflection across the kagome plane $\sigma_h: z \to -z$\cite{li_intertwined_2024,li_origin_2023,friedlan_emergence_2025,schultz_superconductivity_2025}. 
Without loss of essential degrees of freedom, we thus construct a simplified tight-binding model containing 13 out 30 orbitals of the unit cell which exhibit this symmetry property. Further details of this model as well as the calculated spectral functions are presented in Secs.~H and I of the supp.~materials. 

\begin{figure}[H]
    \centering
    \includegraphics[width=13cm]{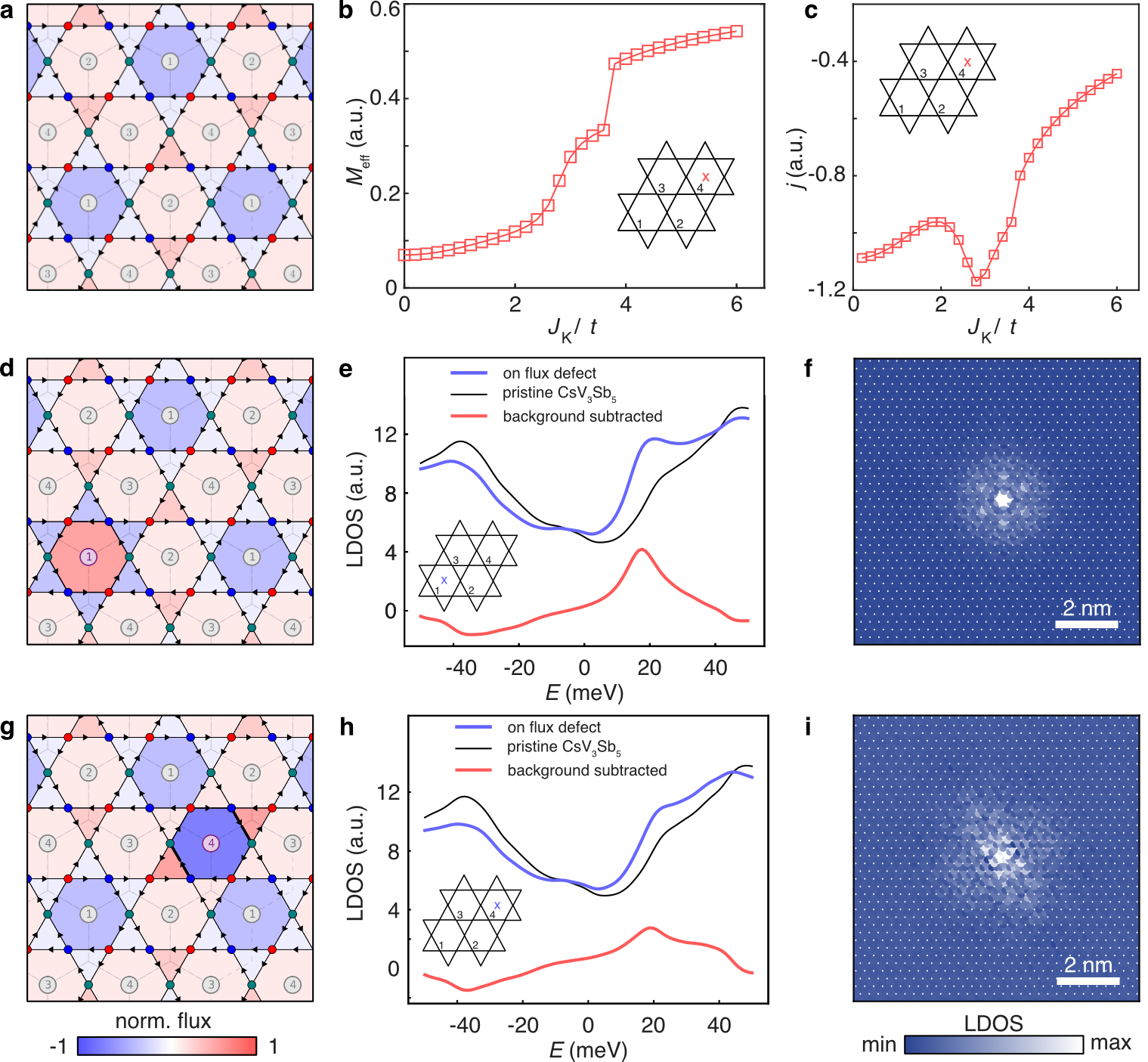}
    \caption{{\bf Theoretical analysis of a flux defect in the loop current order state.} {\bf a}, Shown is a schematic of the kagome lattice, where the atoms of the different sublattice sites are color-coded (blue, green, red). The direction and amplitude of the loop currents between the lattice sites are indicated by black arrows and line thickness, respectively. The red/blue color code indicates the magnetic flux generated by the loop currents in the hexagon and triangle sites where red (blue) indicate flux pointing out of (into) the surface. {\bf b}, shown is the effective hexagon spin moment $M_{eff}$ created by a local bond spin singlet creation at the hexagon site of the magnetic impurity state as a function of the Kondo coupling strength $J_{\bf K}$ normalized by the electron hopping $t$. The impurity position is indicated by a red cross in the inset. {\bf c}, shown is the amplitude of the loop current order $j$ as a function of $J_{\rm K}/t$ at hexagon 4 at the impurity site (see inset). Details of the exact diagonalization calculation are presented in Sec.~F of the supp.~materials. {\bf d,} same as panel a but with the inclusion of a flux defect at hexagon 1. Note the change in the loop current amplitude indicated by the line thickness. {\bf e,} shown is the calculated energy ($E$) dependent local density of states (LDOS) of the pristine sample (black line) and in the presence of a flux defect on hexagon 1 (blue line, see inset). The difference spectrum is shown in red color. {\bf f,} shown is the real space LDOS distribution of the in-gap state of panel e. {\bf g,} same as panel a but with the inclusion of a flux defect at hexagon 4. Note the change in the loop current amplitude and directions indicated by the line thickness and arrows, respectively. {\bf h,} shown is LDOS($E$) of the pristine sample (black line) and in the presence of a flux defect on hexagon 4 (blue line, see inset). The difference spectrum is shown in red color. {\bf i,} shown is the real space LDOS distribution of the in-gap state of panel h.}
    \label{fig:fig4}
\end{figure}

We introduce LCO into our realistic tight-binding model via a mean-field order parameter $\Phi_{\text{MF}}$, which encodes the current pattern on the $d_{zx},d_{yz}$ orbitals on the vanadium sites of the kagome lattice. We focus on the pattern in space group irreducible representation (irrep) $mM_2^+$~\cite{christensen_loop_2022,holbaek_interplay_2025,feng_low-energy_2021}, which has equal weight current on all bonds and is invariant under the inversion symmetry operation. The loop current pattern shown in Fig.~\ref{fig:fig4}(a), which breaks translation symmetry to have a $2a\times 2a$ unit cell, has three ordering wave vectors $J^i_{\mathbf{M}_i}$, with one at each of the three $M$-points ($i=A,\,B,\,C$) in the Brillouin zone. Note that this is the same loop current pattern used for our ED analysis, but here it is generalized to be in terms of vanadium $d_{zx},d_{yz}$ orbitals. We find that the inclusion of $\Phi_{\text{MF}}$ into our model reconstructs the calculated low-energy electronic structure. As we increase the order parameter amplitude, the calculated local density of states (LDOS) develops a pronounced V-shaped dip near the Fermi energy at $\Phi_{\text{MF}} = 0.6$ (measured in units of the hopping), as seen in Fig~\ref{fig:fig4}~(e),(h). This result is in excellent agreement with the $dI/dV$ spectrum recorded on the surface of pristine CsV$_3$Sb$_5$, shown in Fig.~\ref{fig:fig1}g.

We incorporate the local flux defect into our tight-binding model by adding an impurity Hamiltonian $H_{\text{imp}} = V_{\text{imp}} J_{\text{imp}}$, where $V_{\text{imp}}$ quantifies the strength of the flux defect at the defect site and $J_{\text{imp}}$ is the current operator making a loop around a single hexagon. Two distinct scenarios exist, one where the flux defect is added to the $C_{6z}$-symmetric hexagon 1 and another where defect is added to the $C_{6z}$-breaking hexagon 4, as seen in Fig.~\ref{fig:fig4}, d and g. In both cases, we tune $V_{\text{imp}}$ through the relevant hexagon, and study the appearance of a bound state by looking for a pole in the $T$-matrix. Details of these calculations are shown in Sec.~J of the supp.~materials. As seen in Figs.~\ref{fig:fig4}, e and h, our model calculations show that in both cases an additional LDOS peak emerges near the positive-energy edge of the V-shaped gap if $V_{\text{imp}}$ is sufficiently large. The difference spectrum (red solid line) between the LDOS at the defect site and on the pristine surface highlights the spectral characteristics of this in-gap state. Critically, we can verify the bound state character of this spectral feature through a $T$-matrix analysis: the detection of zero crossings of $\text{Re}[T^{-1}(\omega)]$, and simultaneously a small imaginary part in $\text{Im}[T^{-1}(\omega)]$ signify a bound state resonance. Hence, the emergence of an LDOS peak near the positive-energy edge of the V-shaped gap in the presence of a flux defect in the LCO phase is in excellent agreement with our experimental observations of a Co-atom induced $dI/dV$ peak within the V-shaped CDW gap at $V_{\rm B}>0$, shown in Fig.~\ref{fig:fig3}d and Sec.~D.1 of the supp.~materials.

Furthermore, we plot the real-space resolved spectral weight of the LDOS at the bound state energy ($E\approx 20$ meV) in Fig.~\ref{fig:fig4}, f and i. The real-space spectral weight distribution of this bound state is $C_{6z}$-symmetric and $C_{6z}$-breaking for a flux defect on hexagon 1 and 4, respectively. These theoretical results reproduce the experimentally observed point-like and linear $dI/dV$ patterns, respectively, shown in Fig.~\ref{fig:fig3}. We wish, however, to point out that these symmetry properties are a consequence of the underlying symmetry of the local CDW electronic structure (see Sec.~D.2 of the suppl.~materials for discussion)~\cite{christensen_loop_2022}, rather than a defining feature of the LCO state. Finally, we also examine the impact of such a flux defect on the LCO by reconstructing the loop current and flux patterns within the $mM_2^+$ irrep, as shown in Fig.~\ref{fig:fig4},~d and g (see Sec.~I of the suppl.~materials for details)~\cite{christensen_loop_2022,holbaek_interplay_2025,feng_low-energy_2021}. As can be seen, the presence of a flux defect on hexagons 1 and 4 is accommodated by a corresponding change in the loop current amplitude and directions, and the flux defects inherits the $C_{6z}$-preserving and breaking symmetry characteristics of the electronic CDW structure. Critically, the construction of the flux and loop current patterns by a set of three current unidirectional operators guarantees Kirchhoff's law: no charge accumulates locally, even in the presence of a flux defect.

The results of torque magnetometry and muon spin relaxation measurements suggest an upper bound $m_{\rm LC}\leq0.02\,\mu_{\rm B}$ for the magnetic moment $m_{\rm LC}$ of the loop current order ($\mu_{\rm B}$ denotes the Bohr magneton)~\cite{kenney2021absence, gui2025probing}. Given a local flux defect, as seen in Fig.~\ref{fig:fig4}, d and g, carries a comparable additional moment, we would expect a negligible Zeeman shift $E_{\rm z}/B$ of this $dI/dV$ peak on the order of $E_{\rm z}/B\leq1\,\mu$eV/T when an external magnetic field $B$ is applied along the out-of-plane direction. We tested the response of this low-energy $dI/dV$ peak to a magnetic field $B=\pm1\,$T applied along the out-of-plane direction, as shown in Sec.~K of the supp.~materials. In agreement with this estimate, we cannot detect any field-induced spectral changes to this in-gap state at $V_{\rm B}\approx10\,$mV. Note that we are also unable to detect a $B$-induced change in the chirality of the CDW, as previously reported~\cite{jiang2021unconventional, xing2024optical}, possibly owing to the presence of the Co atoms that pin the CDW domains~\cite{jiang2021unconventional, nakazawa2025origin}.

\section{Conclusion}
Our work provides spectroscopic insight into the quasiparticle excitations of the $2a\times2a$ CDW in CsV$_{3}$Sb$_{5}$, in particular with respect to the hidden order in the CDW state at low temperatures. Results from temperature- and magnetic-field dependent STM measurements as well as control experiments show that the magnetic moment of surface-adsorbed Co atoms induces a low-energy $dI/dV$ peak inside the V-shaped gap of the $2a\times2a$ CDW at $T<30\,$K. Comparing our observations with results from quantum many-body simulations on the microscopic Kondo coupling between a magnetic impurity and electronic orbital magnetism as well as realistic tight-binding model calculations, we interpret this spectral feature as the quasiparticle excitation of a loop-current order state that exists in the CDW phase at $T<30\,$K. Therefore, our study provides a microscopic picture for the observation of TRS-breaking orbital magnetism in the out-of-plane direction and anomalous transport detected in macroscopic measurements of bulk samples~\cite{guo2022switchable, khasanov2022time, gui2025probing, wang2025long}. We discuss limitations of our study and analyses in Sec.~L of the suppl.~materials.

From a theoretical viewpoint, LCO and CBO are expected to appear in concert, owing to a symmetry-allowed cubic coupling term between the two order parameters~\cite{park2021electronic}. Given the onset of the $2a\times2a$ CDW phase at $\approx90\,$K~\cite{li2022discovery}, it is thus conceivable that a preexisting CBO would enhance the tendency toward the formation of an LCO phase at $T\approx30\,$K via the same cubic coupling term, suggesting that CBO is the primary ordered phase with higher energy scale~\cite{alkorta2026symmetry}. Looking ahead, it will also be interesting to examine the nature of the superconducting phase appearing at $T<3\,$K~\cite{ortiz2020cs} and the CDW phase at $T>30\,$K using this method based on single atom magnets. More broadly, the local Kondo coupling between a magnetic moment and the electrons of a system with spontaneous time-reversal symmetry breaking could provide new avenues for examining the nature of other interesting quantum phases of matter, such as orbital insulators and fractional Chern states of moiré and graphene superlattices \cite{lu2019superconductors, wu2019topological, ledwith2020fractional, abouelkomsan2020particle, li2021spontaneous, devakul2021magic,park2023observation, xu2023observation, lu2024fractional, luThermodynamic2024, luFractionalQuantum2024,song2024density}.

\bibliography{bibliography}

\clearpage
\section{Figures}

\newpage
\section{Extended Data Figures}
\setcounter{figure}{0}
\makeatletter 
\renewcommand{\figurename}{Extended Data Figure}
\makeatother

\begin{figure}[H]
    \centering
    \includegraphics[width=1\linewidth]{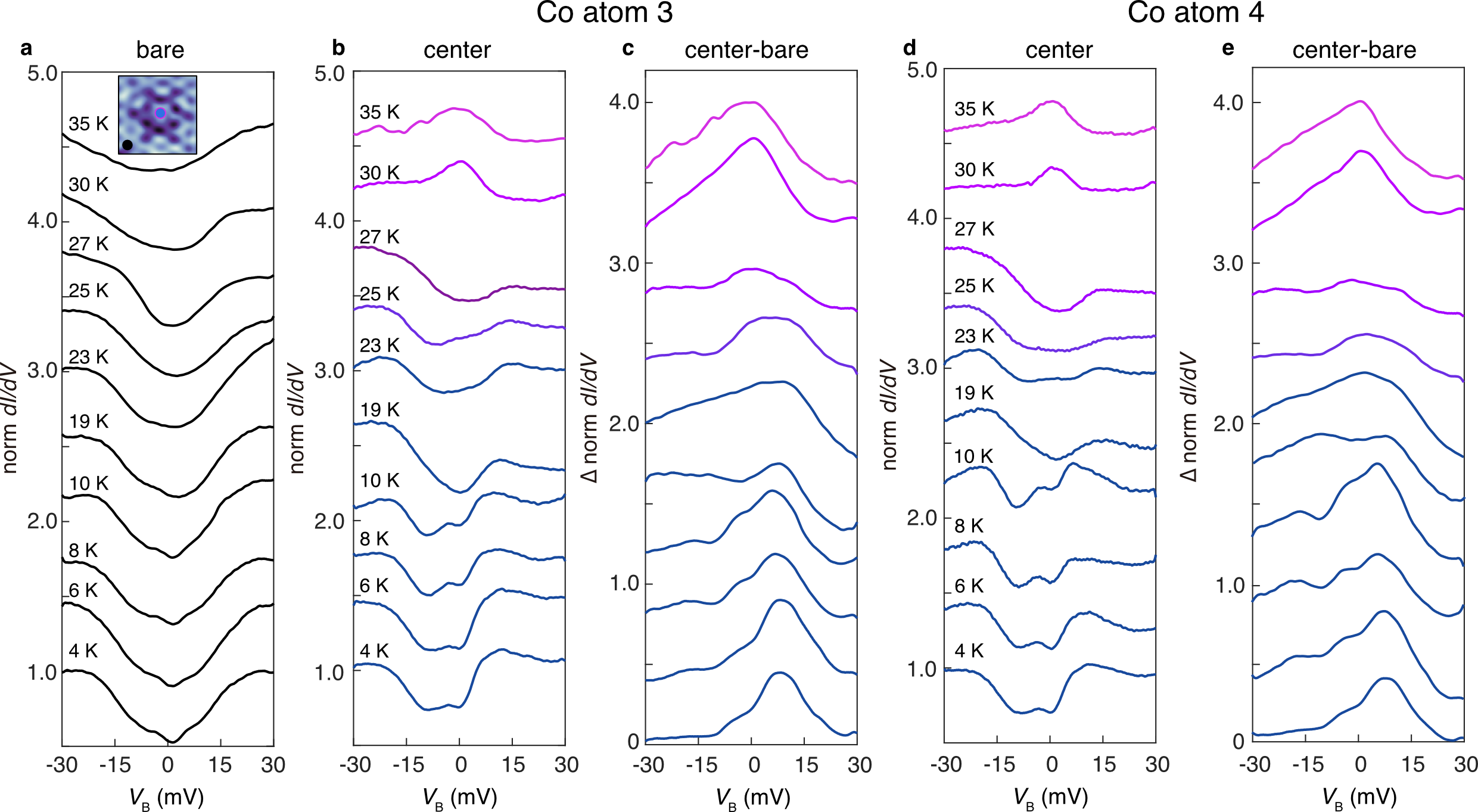}
    \caption{{\bf Temperature evolution of the $dI/dV$ peak at $V_{\rm B}\approx10\,$mV induced by Co atoms deposited on pristine CsV$_3$Sb$_5$.} {\bf a,} Shown are normalized $dI/dV$ spectra recorded on the bare substrate away from Co atoms at different indicated temperatures. The tip position is schematically indicated by a black dot in the inset. {\bf b,} Shown are normalized $dI/dV$ spectra recorded at the center of the Co atom 3 at different indicated temperatures. The tip position is schematically indicated by a blue dot in the inset of panel a. {\bf c,} Shown are the $dI/dV$ spectra of panel b after subtracting the background $dI/dV$ spectra of panel a at the corresponding temperatures. {\bf d,} Shown are normalized $dI/dV$ spectra recorded at the center of the Co atom 4 at different indicated temperatures. {\bf e,} Shown are the $dI/dV$ spectra of panel d after subtracting the background $dI/dV$ spectra of panel a at the corresponding temperatures.  All spectra were recorded with $V_{\rm B}=-30\,$mV, $I=3\,$nA, and $V_{\rm m}=1\,$mV and are vertically offset for clarity.}
    \label{fig:extfig2}
\end{figure}

\begin{figure}[H]
    \centering
    \includegraphics[width=1\linewidth]{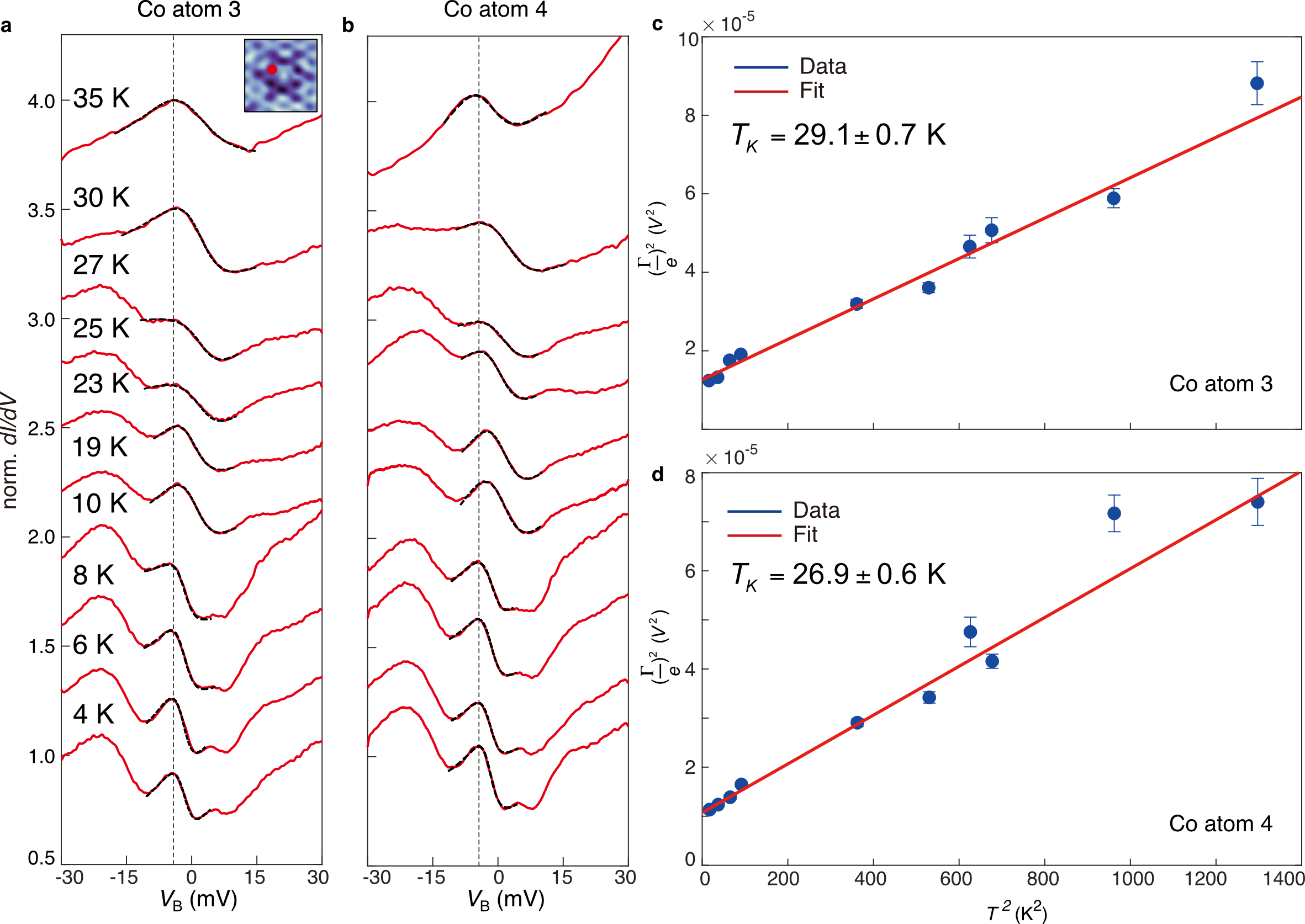}
    \caption{\justifying{\bf Temperature evolution of the Kondo effect of Co atoms on pristine CsV$_3$Sb$_5$.} {\bf a} and {\bf b,} shown are normalized $dI/dV$ spectra recorded at different indicated temperatures (solid lines) and corresponding fits (dashed lines) for two different Co atoms labeled '3' and '4' deposited on the surface of CsV$_3$Sb$_5$ ($V_{\rm B}=-30\,$mV, $I=3\,$nA, $V_{\rm m}=1\,$mV). The spectra were recorded one unit cell away from the Co atom position, as shown in the inset of panel a, and vertically offset for clarity. The vertical black dashed lines delineate the spectral evolution of the Kondo $dI/dV$ peak. Details of the Kondo fitting procedure are described in Sec.~C.2 of the supp.~materials. {\bf c} and {\bf d,} Shown is the dependence of the Fano function linewidth $\Gamma$ on the temperature $T$ as extracted from least-square fits to the temperature dependent $dI/dV$ spectra of Co atoms 3 and 4, shown in panels a and b. We plot $\Gamma^2(T^2)$ so that that indicated Kondo temperature $T_{\rm K}$ can be directly determined from a finite vertical intercept at $T^2=0$.}
    \label{fig:extfig1}
\end{figure}

\newpage
\section{Methods}
\subsection{Synthesis of (Ti-doped) CsV$_3$Sb$_5$ crystals}

Single crystals of CsV$_3$Sb$_5$ and Cs(V$_{0.95}$Ti$_{0.05}$)$_3$Sb$_5$ were grown via a conventional flux-based growth technique. Vanadium powder purchased from the Sigma-Aldrich was cleaned using a mixture of isopropyl alcohol and Hydrochloric acid to remove residual oxides. Cs (liquid, Alfa 99.98\%), V (powder, Sigma 99.9\%), Ti (powder, Alfa 99.9\%), and Sb (shot, Alfa 99.999\%)  were loaded inside a milling vial with the required stoichiometries and then sealed in an Argon-atmosphere. To grow the single crystals of Cs(V$_{1-x}$Ti$_x$)$_3$Sb$_5$ ($x$=0, 0.05), elemental stoichiometries, respectively, taken as Cs$_{20}$V$_{15-x}$Ti$_x$Sb$_{120}$ ($x$=0, 3) and then milled for about an hour. After milling, the powders were poured into the alumina crucibles and sealed inside a stainless steel tube. The samples were heated at 1000$^\circ$ C for 10 h and then cooled to 900$^\circ$ C at 25$^\circ$ C/hr. Below 900$^\circ$ C, the samples were cooled at 1$^\circ$ C/hr to 500$^\circ$ C. Once the growth period was over, the plate-like single crystals were separated gently from the flux and then cleaned with ethanol.  

\subsection{Scanning Tunneling Microscopy (STM) Measurements}

The undoped and titanium-doped CsV$_{3}$Sb$_5$ samples were cleaved after cooling down to a temperature $T=4\,$K inside an ultra-high vacuum (UHV) chamber with a base pressure of $p\approx1.4 \times 10^{-10}\,$mbar. Several crystals of doped and undoped CsV$_{3}$Sb$_5$ were cleaved and the results presented in this manuscript were consistently observed. An electron-beam evaporator (Focus GmbH) was used to deposit cobalt and vanadium atoms from high-purity wire (99.99\%) onto the samples held at $T=4\,$K inside the scanning tunneling microscope (STM). STM measurements were conducted using a home-built STM instrument under cryogenic (4\,K$\leq T\leq35\,$K) and UHV ($p\approx1.4 \times 10^{-10}\,$mbar) conditions using a chemically etched tungsten tip. The tip was prepared on a Cu(111) surface through field emission and controlled indentation, as well as calibrated against the Cu(111) Shockley surface state before each set of measurements. Bias voltage ($V$) dependent differential conductance ($dI/dV$) spectra and maps were recorded using standard lock-in methods with a bias modulation $1\,\text{mV}\leq V_{\rm m}\leq10\,\text{mV}$ at a frequency $f=3.971\,$kHz, as indicated in the main text. The $dI/dV$ maps were recorded using multi-pass mode to avoid set-point effects. To enhance the clarity of the data presentation, high-frequency noise originating from mechanical vibrations coupling to the tip-sample junction was removed from the raw data and interpolation was performed to reduce pixelated appearance where applicable. \\

\section{Acknowledgments}

We gratefully acknowledge valuable discussions with Philip Moll, Xi Dai, Jia-Xin Yin, and Bent Weber. This work was primarily supported by the Hong Kong RGC (Grant Nos. 26304221, 16302422, 16302624, 16304525, and C6033-22G) and the Croucher Foundation (Grant No. CIA22SC02) awarded to B.J and Z.Y.M (HKU C7037- 22GF). Z.Y.M. acknowledges support by the Hong Kong RGC (Grant Nos. AoE/P701/20, 17309822, 17302223), the ANR/RGC Joint Research Scheme sponsored by the Hong Kong RGC and the French National Research Agency (Project No. A HKU703/22). G.P. acknowledges support from the Würzburg-Dresden Cluster
of Excellence on Complexity and Topology in Quantum Matter - ct.qmat (EXC 2147, Project No. 390858490). S.D.W., G.P., and A.C.S. gratefully acknowledge support via the UC Santa Barbara NSF Quantum Foundry funded via the Q-AMASE-i program under award DMR-1906325. D.P. acknowledges support by the National Natural Science Foundation of China through the Excellent Young Scientists Fund (Grant No. 22022310). H.C.P. acknowledge support of the National Key R\&D Program of China (Grants No.2021YFA1401500), the Hong Kong RGC (Grant No. 26308021), and the Croucher Foundation (Grant No. CF21SC01). C.Y.C. acknowledges support from the Tin Ka Ping Foundation. H.Q.W. acknowledges support from NSFC (Grant No. 12474248) and Guangdong Fundamental Research Center for Magnetoelectric Physics (Grant No.2024B0303390001). This work was supported by the German Research Foundation (DFG) through CRC TRR 288 ``Elasto-Q-Mat,'' project A07 (D.J.S. and J.S.) and the Simons Foundation Collaboration on New Frontiers in Superconductivity (Grant SFI-MPS-NFS-00006741-03) (J.S.).

\section{Author Contributions}

B.J., C.Y.C., and J.Z. conceived the project. J.Z. and C.Y.C. conducted the scanning tunnelling microscopy measurements and analyzed the experimental data. G.P., X.Z., Y.D.L., Z.Y.M., and B.J. designed the theoretical interpretation. D.J.S. and J.S. performed the tight-binding and loop current pattern calculations. G.P., A.C.S., and S.D.W. synthesized the samples used in the study. C.C. and D.P. performed density functional theory calculations. H.Q.W. performed the exact diagonalization calculations. All authors discussed the results and contributed to the writing of the manuscript.

\section{Competing Interest Declaration} The authors declare no competing financial interest.

\section{Data Availability Statement} Replication data for this study can be accessed on Zenodo via link XXX.

\end{document}